\documentclass[aps,pra,twocolumn,showpacs,10pt,superscriptaddress,a4paper,longbibliography]{revtex4-1}
\usepackage{graphicx}
\usepackage{amssymb,amsmath}
\usepackage{epsfig}
\usepackage{txfonts}

\usepackage{amsmath,amssymb,amsfonts}
\usepackage{mathrsfs}
\usepackage{empheq}

\usepackage{color}
\usepackage[unicode,breaklinks]{hyperref}

\usepackage[unicode]{hyperref}
\hypersetup{
	pdftitle={cluster},
	pdfauthor={Yu},
	pdfsubject={Vortex},
	colorlinks=true,
	linkcolor=blue,
	citecolor=blue,
	filecolor=black,
	urlcolor=blue
}

\newcommand{\be}{\begin{equation}}
\newcommand{\ee}{\end{equation}}
\newcommand{\bea}[1]{\begin{align}#1\end{align}}

\newcommand{\nn}{\nonumber }

\begin{document}


\title{Axis-symmetric Onsager Clustered States of  Point Vortices in a Bounded Domain }


%
%

\author{Yanqi Xiong}
\affiliation{Graduate School of China Academy of Engineering Physics, Beijing 100193, China}
\author{Jiawen Chen}
\affiliation{Graduate School of China Academy of Engineering Physics, Beijing 100193, China}
\author{Xiaoquan Yu}
\email{xqyu@gscaep.ac.cn}
\affiliation{Graduate School of China Academy of Engineering Physics, Beijing 100193, China}
\affiliation{Department of Physics, Centre for Quantum Science, and Dodd-Walls Centre for Photonic and Quantum Technologies, University of Otago, Dunedin, New Zealand. } 

%
\date{\today}

\begin{abstract}
We study axis-symmetric Onsager clustered states of  a neutral point vortex system confined to a  two-dimensional  disc.   Our analysis is based on the mean field of bounded point vortices in the microcanonical ensemble.  The clustered vortex  states are specified by the inverse temperature $\beta$  and the rotation frequency $\omega$,  which are the conjugate variables of energy $E$ and angular momentum $L$, respectively.  The formation of  the axis-symmetric clustered vortex   states (azimuthal angle independent)   involves the  separating of vortices with opposite circulation and  the  clustering of vortices with same circulation  around  origin and  edge.  The state preserves $\rm SO(2)$ symmetry while breaks $\mathbb Z_2$ symmetry.  We find that,  near the uniform state ($E=0$), the  rotation free state ($\omega=0$) emerges at particular values of  $L^2/E$ and  $\beta$.   At large energies,  we obtain asymptotically exact vortex density distributions,  whose validity  condition  gives rise the lower bound of $\beta$ for the rotation free states.  Noticeably,  the  obtained vortex density distribution near the edge at large energies provides a novel exact vortex density distribution for the corresponding chiral vortex system.
 
  
\end{abstract}

\maketitle
\section{Introduction}
In two-dimensional (2D) fluid turbulence,  energy at small scales can transport to large scales  known as inverse energy cascade~\cite{EyinkRMP, kraichnan1980two,  TABELING20021, Boffetta2012}. This process involves formations of large scale vortex patterns.  Onsager explained the formation of large scale structures by studying  equilibrium statistical mechanics of point vortices in a bounded domain.  The macroscopic  vortex structure is associated with clustering of  like-sign point vortices at negative temperature~\cite{Onsager1949, EyinkRMP}. These coherent large structures occur  in various  systems. Examples are Great Red Spot in Jupiter's atmosphere~\cite{GreatRedSpot}, giant vortex clusters in atomic Bose-Einstein condensates (BECs)~\cite{Gauthier2019, Johnstone2019, Mattprx}, and vortex clustering in quantum fluids of exciton–polaritons~\cite{panico2023onset}. 

Clustering phenomena of vortices have attracted much attention~\cite{MJ74, edwards1974negative, williamson1977, Pointin, campbell1991statistics, SmithPRL, SmithONeil, Yatsuyanagi2005, esler2015, DecayingQTBillam,  siggia1981, IECMatt, Matt2012, TomPRA, TapioPRL, Gurarie2004, clusteringYu,  HayderPRA2016, Valani2018, Junsik2018, Junsik2019, AnghelutaPRE,  AudunPRE2017, Guo2021, Vishal2021, review2017, chavanis1996, Chavanis2002}.  For a given 2D domain,  searching for the maximum entropy  clustered vortex  state is at the center of investigations.   For circularly symmetric domains,   previous studies on neutral vortex systems  focus on zero angular momentum case~\cite{Yatsuyanagi2005,clusteringYu,TapioPRL, AnghelutaPRE, AudunPRE2017}.  The role of finite angular momentum in formation of  clustered states in a neutral vortex system remains less well-explored. 

In this paper,  we  study axis-symmetric clustered vortex states through the mean field approach. The mean field theory to describe formation of  negative temperature  clustered vortex states  was formulated systematically  by Joyce and Montgomery~\cite{JM73}. The mean field equations, which were obtained via maximizing the entropy of the vortex system,  are essential for analyzing possible clustered states.   We consider  a neutral vortex system consisting of equal number of positive and negative vortices confined to a disc.  For given positive vortex number $N_{+}$ and negative vortex number $N_{-}$,  clustered vortex states are specified by energy $E$ and angular momentum $L$ or their conjugate variables inverse temperature $\beta (E,L)$ and  rotation frequency $\omega(E, L)$.  We find that  in the limit $\beta \rightarrow 0$, $\omega \rightarrow \infty$ while keeping $\beta \omega$ finite,  positive  and negative vortex density distributions are Gaussian distributions centered at  origin and edge, respectively. For rotation free sates ($\omega=0$),  we find asymptotically  exact  positive and negative vortex density distributions at large  energies.   In particular,  the one maximized on the  edge provides a new exact solution to the  mean field equations for the corresponding chiral vortex system.  The lower bound of $\beta$ is obtained from the validity condition of  the asymptotically  exact  solutions at high energies, above which rotation free clustered states exist.   
To analyze clustered states closed to the uniform state at low energies,  we generalized the perturbation theory, which was initially developed for chiral systems~\cite{SmithONeil}, to the neutral case.   Using this perturbation theory we find the critical value of $\beta$ for the onset of  the rotation free  clustered vortex state, providing an upper bound of $\beta$.

\section{Model} 
The point-vortex model describes dynamics of  well-separated quantum vortices in a superfluid at  low temperature~\cite{AHNS1980},   2D classical inviscid, incompressible fluids~\cite{Aref1999, siggia1981} and  guiding-center plasma~\cite{JM73}. Negative temperature states occur due to the bounded phase space of a 2D confined point vortex system.  Above a certain energy the number of available states decreases as the function of energy and consequently the system becomes more ordered as energy increases~\cite{EyinkRMP} .

We consider a  system consisting a  large  number of  point  vortices  confined to a uniform disc of radius $R$.  The system is neutral and  contains  $N_+$ positive
vortices and $N_-=N_+$ negative vortices . The  Hamiltonian  is~\cite{newton2001n}
\bea{
	\label{Hamiltonian1}
	H &= -\sum_{i\neq j} \kappa_i \kappa_j \log|{\boldsymbol r}_i-{\boldsymbol r}_j|+ \sum_{i,j}\kappa_i \kappa_j \log\bigg|({\boldsymbol r}_i-\bar{\boldsymbol r}_j)\frac{|{\boldsymbol r}_j|}{R}\bigg|.
}
In a BEC,  the Hamiltonian Eq.~\eqref{Hamiltonian1} is measured in unit $E_0=\rho m_a \kappa^2/4\pi$, 
where $\rho$ is the superfluid density, $\kappa=h/m_a$ is the circulation quantum and $m_a$ is the atomic mass.   In this unit,  $\kappa_{i}=\pm 1/N_{\pm}$ and the $1/N_{\pm}$ scaling gives a well-defined mean field limit~\cite{eyink1993negative, caglioti1995special}.  
For a vortex at position ${\boldsymbol  r}_j$,  its image locates at 
$\bar{{\boldsymbol r}}_j=R^2{\boldsymbol r}_j/|{\boldsymbol  r}_j|^2$ to ensure that the fluid velocity normal to the boundary vanishes.  The Hamiltonian \eqref{Hamiltonian1} has rotational $\rm SO(2)$ symmetry due to the disc geometry and $\mathbb Z_2$ symmetry (invariant under $\kappa_i \rightarrow -\kappa_i$).  Hereafter we set $R=1$ without loosing generality.

To investigate  formations of large-scale  clustered patterns, it is necessary to consider  
the continuous effective Hamiltonian in the large $N$ limit~\cite{JM73}:
\bea{
\label{Hamiltonian2}
\ H_{\rm eff}=\frac{1}{2} \int {\rm d}^2 {\boldsymbol r} {\rm d}^2{\boldsymbol r}' \, \sigma({\boldsymbol r})\phi({{\boldsymbol r}-{\boldsymbol r}'})\sigma({\boldsymbol r}').
}
Here  $\sigma (\boldsymbol{r})\equiv n_+(\boldsymbol{r})-n_-(\boldsymbol{r})$
is the vorticity field, 
\bea{n_{\pm}(\boldsymbol {r})\equiv \frac{1}{N_{\pm}}\sum_{i}\delta(\boldsymbol{r}-\boldsymbol{r}^{\pm}_{i})}
is the local density of positive (negative) vortices, and $\boldsymbol{r}^{\pm}_i$ is the position of the vortex $i$ with circulation $\pm 1/N_{\pm}$. The vortex densities $n_{\pm}$ satisfy the normalization condition
\bea{
\label{normalization}
\int {\rm d}^2 \boldsymbol {r} \, n_{\pm} =1.
}
The Green's function $\phi(\boldsymbol{r}-\boldsymbol{r}')$ satisfies $\nabla^2 \phi(\boldsymbol{r} -\boldsymbol{r}')=-4\pi \delta (\boldsymbol{r} -\boldsymbol{r}')$. Here $\phi (\boldsymbol {r}-\boldsymbol {r}')=0$ on the boundary
($|\boldsymbol {r}|=1$), and $\phi (\boldsymbol{r}-\boldsymbol {r}')\sim -2 \protect \qopname \relax o{log}|\boldsymbol {r}-\boldsymbol {r}'|$ as $|\boldsymbol{r}-\boldsymbol {r}'|\rightarrow
0$~\cite {Lin1941}. The stream function 
\bea{
	\label{streamfunction}
	\psi(\boldsymbol{r})\equiv\int {\rm d}^2 \boldsymbol{r}'\phi(\boldsymbol{r},\boldsymbol{r}')\sigma (\boldsymbol{r}'),
}
satisfies the Poisson equation 
\bea{
	\label{Poisson}
	\nabla^2 \psi =- 4 \pi \sigma(\boldsymbol {r})
}
with the boundary condition $\psi(r=1,\theta)=C$. Here $C$ is a constant. Recall that the radial velocity 
\bea{\label{bc} u_r = \frac{1}{r} \frac{\partial \psi}{\partial \theta}.}
This boundary condition ensures that 
there is no flow across the boundary of the domain. 
Without losing generality, we choose $C=0$, which is equivalent to
including image terms in Eq.~\eqref{Hamiltonian1}.

For a rotationally symmetric domain,  energy 
\bea{
E=\frac{1}{2} \int {\rm d}^2 {\boldsymbol r} \, \sigma \psi  
}
and angular momentum 
\bea{
L=\int {\rm d}^2 \boldsymbol{r} \, r^2\sigma}
are conserved quantities.

The most probable density distribution is given by maximizing the entropy function 
\bea{
\label{entropyfunction}
S= - \int {\rm d}^2 \boldsymbol{r} \, n_+ \log n_+ - \int {\rm d}^2 \boldsymbol{r}\, n_-\log n_- ,
}
at  fixed values of $N_+$, $N_-$, $E$ and $L$ ~\cite{JM73}.  From the variational equation   
\bea{\delta S -\beta \delta E- \alpha \delta L- \mu_+ \delta N_+/N_+ -\mu_- \delta N_-/N_- =0,}
we obtain 
\bea{
\label{selfconsistenteq}
n_\pm (\boldsymbol{r}) = \exp \left[\mp \beta \psi(\boldsymbol{r})\mp \alpha r^2 +\gamma_\pm \right],
}
where $\beta$, $\alpha$ and $\mu_\pm$ are Lagrange multipliers and $\gamma_{\pm}=-\mu_{\pm}-1$. 
The parameters $\beta$, $\omega \equiv \alpha/\beta$ and $\mu_{\pm}$ have the interpretation of inverse temperature, rotation frequency and chemical potentials, respectively.

\section{Onset of clustering}
\label{onset}
In this section we analyze the possible stable large scale coherent structures described by Eq.~\eqref{Poisson} and Eq.~\eqref{selfconsistenteq} near the uniform state. 
Here we generalized the method which was developed for analyzing chiral vortex matter~\cite{SmithONeil}, to the neutral case. 

Let us start at a solution $n_{\pm}$ of 
Eq.~\eqref{selfconsistenteq} at energy $E$ and angular momentum $L$, and consider a nearby 
solution $n_\pm + \delta n_\pm$ at $E+\delta E$ and $L+\delta L$. 
The corresponding changes are

\bea {
	\label{nor1}
	0 &= \int {\rm d}^2 {\boldsymbol r} \,  \delta n_+, \\
	\label{nor2}
	0&= \int {\rm d}^2 {\boldsymbol r} \,\delta n_-, \\
	\label{energy}
	\delta E&= \int {\rm d}^2 \boldsymbol{r} \, \psi \delta \sigma + \frac{1}{2}\int {\rm d}^2 \boldsymbol{r} \, \delta\psi \delta \sigma, \\ 
	\delta L&= \int {\rm d}^2  \boldsymbol{r} \, r^2 \delta \sigma \label{angularmomentum}.
}

To leading order, we obtain 
\bea{
	\label{variationdensity}
	\delta n_+ \simeq &  n_+ (- \psi \delta \beta  - \beta \delta \psi + \delta \gamma_+ - r^2 \delta \alpha), \nn\\
	\delta n_- \simeq & n_- (\psi \delta \beta  + \beta \delta \psi + \delta \gamma_-+ r^2 \delta \alpha), 
}
where $\delta \gamma_-,\delta \gamma_+, \delta\beta$ and $\delta \alpha$ are changes of Lagrange multipliers. Plugging Eq.~\eqref{variationdensity} into Eqs.~\eqref{nor1}-\eqref{angularmomentum}, we have  
\be
\label{variation}
{\cal Q}\delta \boldsymbol{\mu}=-\delta \boldsymbol {T} +\beta {\boldsymbol V} \delta\psi,
\ee
where $\delta\boldsymbol{\mu}=(\delta \beta, \delta \gamma_+,\delta \gamma_-,\delta\alpha)^{\rm T}$,
$\delta \boldsymbol{T}=(0,0,\delta L,\delta E)^{\rm T}$, $\boldsymbol{V}\delta \psi=-\int {\rm d}^2 \boldsymbol{r} \, (n_+, n_-, n r^2, n \psi)^{\rm T} \delta \psi$,
\be
{\cal Q} \equiv
\int {\rm d}^2\boldsymbol{r} \left( \begin{array}{cccc}
	n_+ \psi & -n_+ & 0 &  n_+ r^2\\
	n_- \psi & 0 & n_- & n_- r^2\\
	n\psi r^2 & -n_+ r^2 & n_- r^2 & n r^4 \\
	n \psi^2 & -n_+ \psi & n_- \psi & n \psi r^2
\end{array} \right),
\ee
and $n=n_+ + n_-$ is the total density. 
Variation of the Eq.~\eqref{Poisson} gives us
\bea{
	\label{variationPoisson}
	\nabla^2 \delta \psi &= - 4 \pi \left( \delta n_+ - \delta n_- \right)\\
	&= 4 \pi (\psi n \delta \beta + \beta n \delta \psi - n_+ \delta \gamma_+ +  n_- \delta \gamma_- +n r^2 \delta \alpha) .\nn
}

Our aim is to find stable clustered states which emerge from the 
homogeneous state $n_-=n_+=n_0=1/\pi$.  For the homogeneous state, $\sigma=0$, $\psi=0$, $\alpha=0$, $L=0$ and $E=0$.
We assume that $\delta \alpha$ is the same order as $\delta \psi$ and from Eq.~\eqref{variation} we obtain 
\bea{
	0&=\delta \gamma_+ + \delta \gamma_-, \\
	0&=\delta \gamma_+ -\delta \gamma_- -\delta \alpha-2\beta n_0 \int {\rm d}^2 \boldsymbol{r} \, \delta \psi, \\
	\delta L &= \beta n_0 \int {\rm d}^2 \boldsymbol{r} (1-2r^2) \delta \psi -\frac{1}{6} \delta \alpha,\\
		\delta E&=\frac{1}{2}\int {\rm d}^2 \boldsymbol{r} \, \delta\psi \delta \sigma. 
}
Let us introduce operator ${\cal L}$:
\bea{
	\label{zeromode}
	{\cal L} \delta \psi &\equiv \nabla^2 \delta \psi-8\pi n_0 \bigg[\beta \delta \psi -\beta n_0 \int {\rm d}^2 \boldsymbol{r} \delta \psi \\
	&-3\beta n_0 (1-2r^2)\int {\rm d}^2 \boldsymbol{r} (1-2r^2) \delta \psi +3(1-2r^2) \delta L \bigg] 
	= 0. \nn
}
Then Eq.~\eqref{variationPoisson} becomes a zero mode equation of the operator ${\cal L}$.  The onset of large scale vortex clusters occurs if Eq.~\eqref{zeromode} has non-zero solutions. The value of $\beta$ is undefined in the homogeneous phase within our mean field approach and depends on the mode developing from the uniform state. Since the operator ${\cal L}$ is defined on a disc with the Dirichlet boundary condition, it is natural to decompose Eq.~\eqref{zeromode} in azimuthal Fourier harmonics ${\psi_s}$ which is characterized by the mode number $s$ and satisfies $\partial^2{\psi_s}/\partial{\theta^2}=-s^2\psi_s$:
\bea{\label{decomposation}
	\delta \psi =\sum_s \epsilon f_s \psi_s(r,\theta),
}
where $\epsilon \ll 1$ is a small amplitude and $f_s$ is the mode coefficient.  Then each mode satisfies 
\bea{
	\label{zmeq}
	{\cal L} \psi_s(r,\theta)=0,
}
where $\psi_s(r,\theta)$ satisfies the boundary condition $\psi_s(r=1,\theta)=0$.   We denote
$\delta L=L_0 \epsilon$, $\delta E=E_0 \epsilon^2$ and $\delta \alpha=\epsilon \beta\omega$. 

We find that  
\bea{
	\label{zeromodeg}
	\psi_s(r,\theta)
	&= c_s {\rm J}_s(k r) \cos( s \theta)+b_s+a r^2
}
solves Eq.~\eqref{zmeq} with  
\bea{
	\label{rotationcondition}
	a=-\omega=-2c_s\beta n_0 \int {\rm d}^2 \boldsymbol{r} \, {\rm J}_s(k r) \cos( s \theta)
}
and 
\bea{\beta=-\frac{k^2}{8\pi n_0}.} 
Here ${\rm J}_s(r)$ is the Bessel function of the first kind. Consistently, 
\bea{\label{amc}
	L_0&=\beta n_0 \int {\rm d}^2 \boldsymbol{r}\, (1-2r^2)\left[\psi_s(r,\theta)+\omega r^2\right], \\
	\label{ec}
	E_0&= -\frac{1}{8 \pi}\int {\rm d}^2 \boldsymbol{r}\, \psi_s(r,\theta) \nabla^2 \psi_s(r,\theta).
}
For given $L_0$ and $E_0$,  the parameters $c_s$, $k$, and $b_s$ are determined by Eqs.~\eqref{amc} and \eqref{ec} combined with the Dirichlet boundary condition 
\bea{
	\label{boundarycondition}
	\psi_s(r=1,\theta)=c_s {\rm J}_s(k) \cos(s \theta)+b_s+a=0.
}
The single-valueless of the stream function requires that $s$ has to be an integer, namely, $s\in \mathbb Z$. 

For $s \neq 0$, $L_0=0$, $a=-\omega=0$, $b_s=0$,
\bea{c_s^2=\frac{16 E_0}{ k \left[k {\rm J}_{s-1}(k){}^2-2 s {\rm J}_s(k) {\rm J}_{s-1}(k)+k {\rm J}_s(k){}^2\right]},}
and $k=j_{s,m}$, where $j_{s,m}$ is the $m$th zero of the Bessel function of the first kind ${\rm J}_s(r)$.

For $s=0$, 
\bea{\label{frequency}
	a&=-2c_0\beta n_0 \int {\rm d}^2 \boldsymbol{r} \, {\rm J}_0(k r)=\frac{1}{2} c_0 k {\rm J}_1(k),\\
b_0&=-a-c_0 {\rm J}_0(k).}

For given $E_0$ and $L_0$, $c_0$ and $k$ are determined by 
\bea{E_0&=\frac{1}{8} c_0^2 k^2 \left[{\rm J}_0(k){}^2+{\rm J}_1(k){}^2\right], \\
L_0&=-\frac{1}{4} c_0 k {\rm J}_3(k).}
It is useful to introduce
\bea{\Gamma(k)\equiv \frac{(\delta L)^2}{\delta E}=\frac{L^2_0}{E_0}=\frac{{\rm J}^2_3(k)}{2 \left[{\rm J}_0(k){}^2+{\rm J}_1(k){}^2\right] }}
as a control parameter. 

\begin{figure}[!h]
	\includegraphics[width=2.8in]{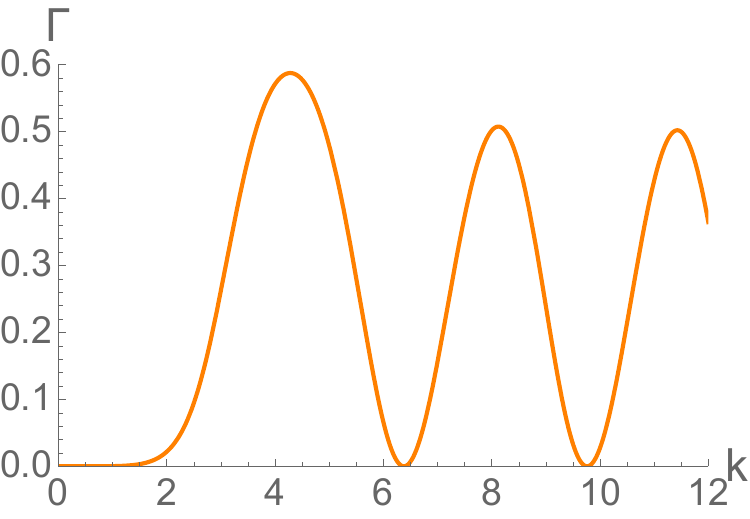}
	\vspace{0.1cm}
	\caption{$\Gamma(k)$ as a function of $k$. The maximum value of  $\Gamma(k)$ is reached at $k=k^{*}$ and $j_{1,1}<k^{*}<j_{2,1}$.
	}
	\label{f:ratio}
\end{figure}

The ratio $\Gamma (k)$ reaches its maximum value at $k=k^{*}$ with $j_{1,1}<k^{*}<j_{2,1}$ (see Fig.\ref{f:ratio}). For a given $\Gamma_0 < \Gamma(k^{*})$, there are more than one values of $k_c$ such that $\Gamma(k_c)=\Gamma_0$. Guided by the maximum entropy principle, the minimal value of $k_c$ corresponds to the equilibrium state.  For $k\rightarrow 0$, $E_0 \rightarrow 0$, $L_0 \rightarrow 0$ and this mode describes the uniform state.

The modes $s \neq 0$ break $\rm SO(2)$ symmetry and the maximum entropy state for given energy is the clustered vortex dipole state which corresponds to the $s=1$ mode~\cite{clusteringYu}. This clustered  vortex dipole state has been recently realized in BEC experiments~\cite{Gauthier2019}. In this paper, we focus on states related to the $s=0$ mode.

\section {Axis-symmetric  Clustered States}
In this section, we present some (asymptotically) exact results on axis-symmetric distributions of  neutral vortex clusters.  For axis-symmetric states, the boundary condition Eq.~\eqref{bc} which is imposed by the most relevant physical condition is fulfilled automatically.   

\subsection{Gaussian vortex states}
Let us firstly consider $\beta \rightarrow 0$. For finite $\omega$, vortex distributions $n_{\pm}$ must be uniform.  However, when $\omega \rightarrow \infty$ simultaneously such that $\alpha=\omega \beta$ is finite, non-trivial distributions can occur. In this special limit, the vortex densities have the profile of Gaussian distribution:   
 \bea{n_{+}(r)&=- \frac{\alpha  }{ \pi  \left[\exp(-\alpha) -1\right]}\exp \left(-\alpha  r^2\right), \\
n_{-}(r)&=\frac{\alpha }{\pi  \left[\exp(\alpha)-1\right]} \exp \left(\alpha  r^2\right),
}
where $\alpha \in (-\infty, \infty)$. 

The corresponding stream function reads 
\bea{\psi(r)&=\frac{\exp (\alpha) \text{Ei}\left(-r^2 \alpha \right)+\text{Ei}\left(r^2 \alpha \right)-2 \left[\exp(\alpha)+1\right]  \log r }{\exp(\alpha)-1} \nn\\
	&-\frac{\exp(\alpha ) \text{Ei}(-\alpha )+\text{Ei}(\alpha )}{\exp(\alpha)-1},
}
where \bea{\text{Ei}(x)=-\int_{-x}^{\infty } \frac{\exp(-t)}{t} \, dt} is the exponential integral function. The stream function satisfies $\psi(r=1)=0$ and  $d\psi/dr|_{r=1}=0$.

\begin{figure}[!h]
	\includegraphics[width=2.8in]{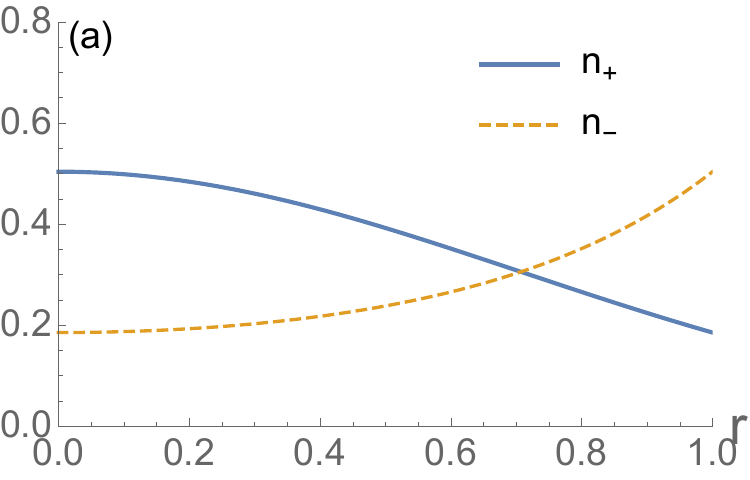}
     \includegraphics[width=2.8in]{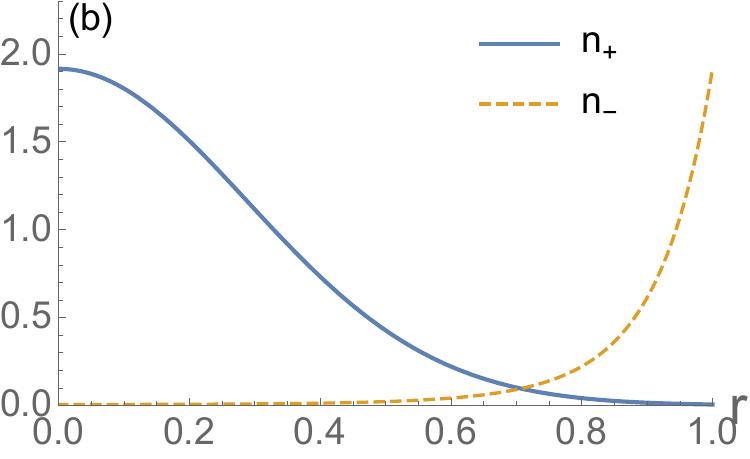}
	\includegraphics[width=2.8in]{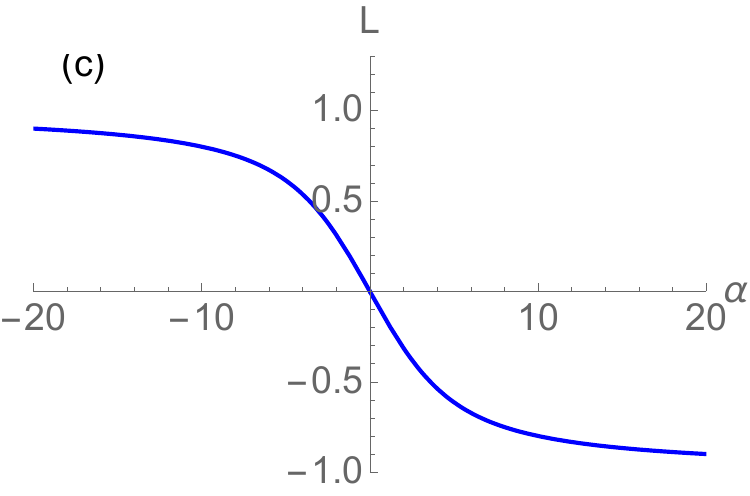}
	\includegraphics[width=2.8in]{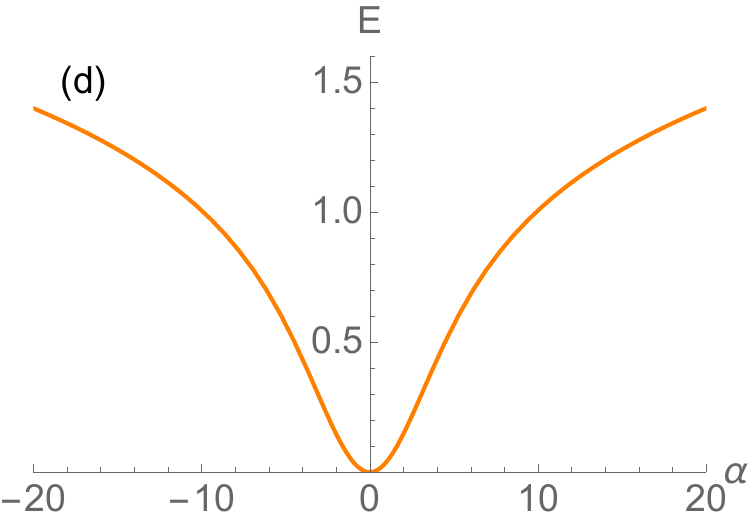}
	\vspace{0.1cm}
	\caption{Vortex densities for $\alpha=1$ (a) and $\alpha=6$ (b). The angular momentum and the energy as functions of $\alpha$ are shown in (c) and (d), respectively.
	}
	\label{f:Gaussian}
\end{figure}

The angular momentum is
\bea{L=\alpha^{-1} \left[2-\alpha  \coth \left(\frac{\alpha }{2}\right)\right].}
It is easy to see that $L \le 1$. Figure~\ref{f:Gaussian} (a)(b) show typical vortex densities for different values of $\alpha$. Figure~\ref{f:Gaussian} (c)(d) show  angular momentum and energy as functions of $\alpha$.  Note that the Gaussian state is available in the chiral vortex system as well~\cite{SmithONeil}.  

\subsection{Rotation free vortex states}

In this subsection, we consider clustered vortex  states for $\omega=0$ and finite $\beta<0$. 

\subsubsection{Onset of axis-symmetric clustered states}
\label{onsetzero}
Closed to the uniform state, the clustered states can be analyzed using the formalism developed in Sec.\ref{onset}.  
The polar angle $\theta$-independent zero modes $s=0$ carry non-zero angular momentum. For $s=0$ modes,
the rotation free condition $a=-\omega=c_0 k {\rm J}_1(k)/2=0$ [see Eq.~\eqref{frequency}]
requires that $k=j_{1,m}$, where $j_{1,m}$ is the $m$th zero of the Bessel function of the first kind ${\rm J}_1(r)$.
These modes occur at 
$\beta=\beta_{1,m}= -j_{1,m}^2/8\pi n_0$ and break  $\mathbb Z_2$ symmetry. The $m=1$ mode starts to emerge at $\beta=\beta_{t}=\beta_{1,1}\simeq -1.835$ and has the highest  
statistical weight among the rotation free modes ($\omega=0$): 
\bea{
	\label{targetzeromode}
	\psi_0(r)=c_0 {\rm J}_{0}(j_{1,1}r)+b_0,
}
where $c_0=\pm 2\sqrt{2}E_0/|{\rm J}_0(j_{1,1})|j_{1,1}$ and $b_0=-c_0 {\rm J}_0(j_{1,1})$. For this mode, $\Gamma (j_{1,1}) \sim 0.545$. Since  $\Gamma (j_{1,1}) >\Gamma (j_{1,m>1})$, $\Gamma (j_{1,1})$ gives the upper bond of $\Gamma$ for the rotation free modes. The rotation free axis-symmetric phase emerges from the uniform phase by varying angular momentum and  energy such that $\Gamma=\Gamma (j_{1,1})$.


\subsubsection{High energy configuration}

All the rotation free and axis-symmetry states satisfy 
\be
\label{symmetricsolution}
\frac{1}{r} \frac{d}{dr} r \frac{d}{dr} \psi(r) = -4 \pi \left[\exp(-\beta \psi(r)+\gamma_+)-\exp(\beta \psi(r)+\gamma_-)\right].
\ee
The most relevant solution of Eq.~\eqref{symmetricsolution} should be the nonlinear continuation of the zero mode $\psi_0$ and describes the axis-symmetry equilibrium state with zero rotation frequency.   

At large energies, vorticies with opposite sign are well-separated and the overlap between $n_{+}$ and $n_{-}$ can be neglected. In this limit,  exact results are available.  Let us assume that positive vorticies are  concentrated in the center of the disc and negative vorticies are distributed along the edge of the disc. The density distribution of positive vortices near $r=0$ can be obtained analytically by neglecting the influence of negative vortices:     
\bea{
	\label{n-positive}
	n_+(r)=\frac{4A}{(2-\pi \beta A r^2)^2}, 
}
with the boundary conditions $\psi(0)=0$ and $\psi'(0)=0$~\cite{SmithONeil}. Here $A=[\pi(1-\beta/\beta_{\ast})]^{-1}$ is fixed by the normalization condition of $n_+$ and $\beta_{\ast}=-2$.  The supercondensation  occurs at $\beta=\beta_{\ast}$,  involving point-like concentration of the positive vortices and the divergence of energy~\cite{kraichnan1975, SmithONeil}.

Near $r=1$, we can neglect the 
influence of positive vortices and find the density distribution of negative vortices
\bea{\label{densitynegative}
	n_-(r)=\frac{2(2/\beta-1)(1-\beta)^2 r^{-2\beta}}{\pi \beta (r^{-2\beta+2}+1-2/\beta)^2},
}
where the boundary conditions are $\psi(1)=0$ and $\psi'(1)=0$.

\begin{figure}[!t]
	\includegraphics[width=2.8in]{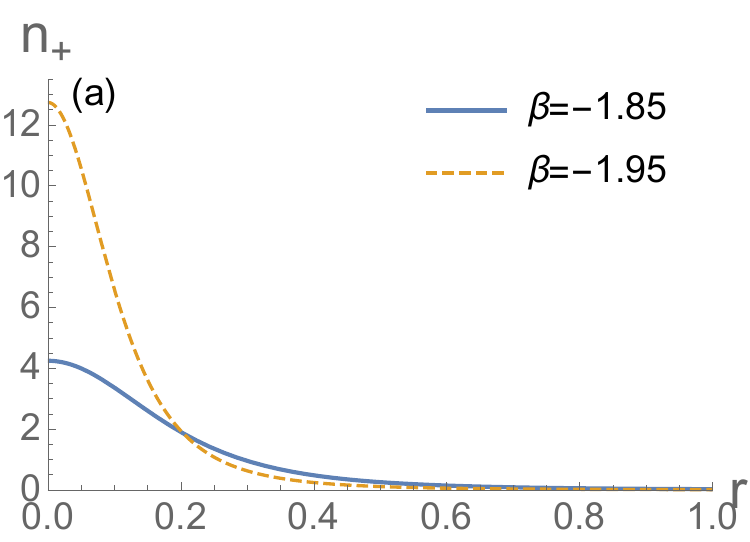}
	\includegraphics[width=2.8in]{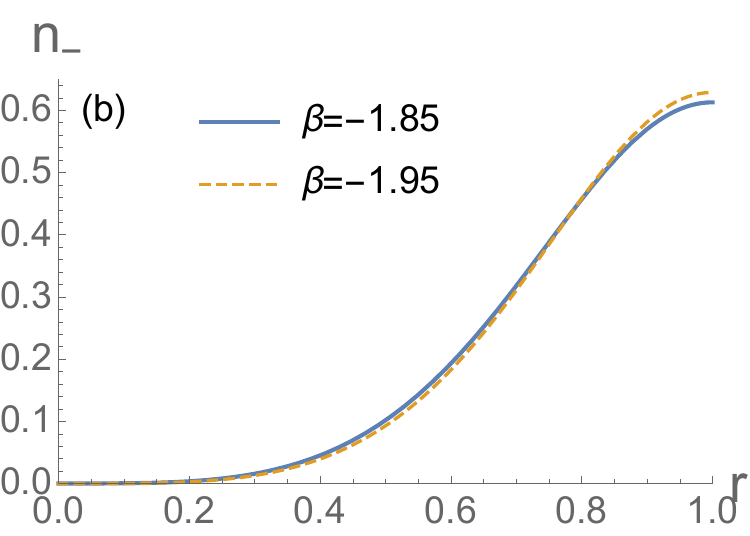}
	\vspace{0.1cm}
	\caption{Vortex density distributions at high energies. The densities  of positive vortices (a) and negative vortices (b) are evaluated via Eq.~\eqref{n-positive} and  Eq.~\eqref{densitynegative}, respectively. 
	}
	\label{f:symmetricphase}
\end{figure}

Note that $\psi(0)$ and $\psi(1)$ can be chosen as arbitrary constants and here we choose them to be zero for convenience. The boundary condition $\psi'(0)=0$ ensures that $n'_+(0)=0$ and $n_{+}$ has no singular behavior near $r=0$. Similarly, the boundary condition $\psi'(1)=0$  implies that $n'_-(1)=0$ and the absence of singular behavior of $n_{-}$ near $r=1$. As approximations of vortex densities at large energies, Eqs.~\eqref{n-positive} and \eqref{densitynegative} should be evaluated for $\beta_{\ast}<\beta$. Combining the critical value of $\beta$ at which the onset of clustering occurs,  we obtain the parameter regime for the rotation free clustered vortex state: 
\bea{\beta_{\ast}<\beta<\beta_t.}
Figure.~\ref{f:symmetricphase} shows the vortex density distributions at high energies.

In the deep clustered sate,  positive vortices are concentrated in a small region and the total energy are contributed dominantly from positive vortices. So as $\beta \rightarrow \beta_{\ast}$,   
\bea{
	E\backsimeq-\frac{2}{\beta^2}\left[\log\left(1-\frac{\beta}{\beta_{\ast}}\right)-\frac{\beta}{2} \right].}

At large energies, the angular momentum is
\bea{
	\label{exactangularmomentum}
	L &=\bigg|\int {\rm d}^2 \boldsymbol{r} r^2 [n_{+}(\boldsymbol{r})-n_{-}(\boldsymbol{r})]\bigg| \nn\\
	&\simeq \bigg|\int {\rm d}^2 \boldsymbol{r} r^2 \bigg[\frac{4A}{\left(2-\pi\beta A r^2 \right)^2} -\frac{2(2/\beta-1)(1-\beta)^2 r^{-2\beta}}{\pi \beta (r^{-2\beta+2}+1-2/\beta)^2} \bigg] \bigg|,
}
and as $\beta \rightarrow \beta_{\ast}$, $L\rightarrow L_{\rm {max}}$ with 
\bea{
	L_{\rm max}=\left|\int {\rm d}^2 \boldsymbol{r} r^2 \left\{\delta(\boldsymbol{r})-18 r^4 [\pi (2+r^6)^2]^{-1} \right\}\right|\simeq 0.705.
}

\section{Exact results for Chiral vortex clusters}
As stated in the previous section, Eq.~\eqref{densitynegative} is the exact solution to Eq.~\eqref{symmetricsolution}, provided  that $n_{+}$ is neglected. Hence Eq.~\eqref{densitynegative} provides an exact vortex density distribution for a chiral system, which is distinct from the well-known exact distribution. In this section, we make a summary of   relevant exact results and make a comparison between our finding and the known distribution.

\begin{figure}[h!t]
	\includegraphics[width=2.8in]{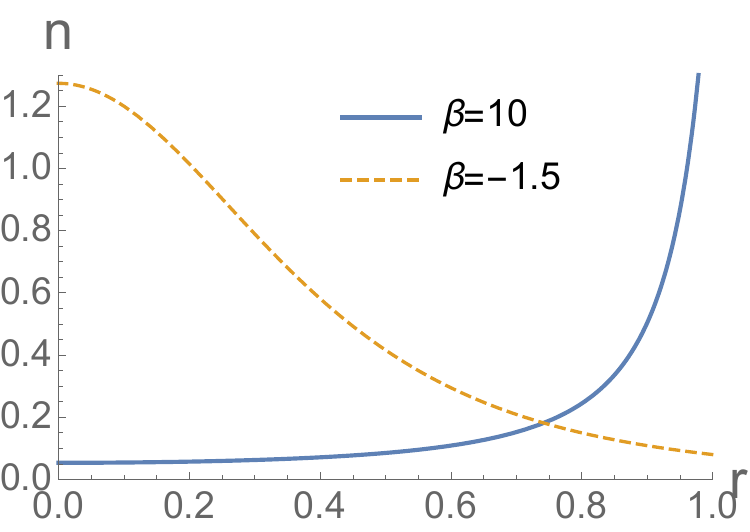}
	\vspace{0.1cm}
	\caption{Typical profiles of the vortex density distribution described by Eq.~\eqref{exactdensity0} in two distinct parameter regimes: $0<\beta$ and $-2<\beta<0$.  
	}
	\label{f:exact0}
\end{figure}

For a rotation free ($\omega=0$) and axis-symmetric chiral system, Eq.~\eqref{symmetricsolution} becomes
\bea{
\label{chiral1}
\frac{1}{r} \frac{d}{dr} r \frac{d}{dr} \psi(r) &= -4 \pi n(r), \\
\label{chiral2}
n(r)&=\exp(-\beta \psi +\gamma).
}
There is a known exact solution to Eqs.~\eqref{chiral1} and~\eqref{chiral2}, which is Eq.~\eqref{n-positive}:
\bea{\label{exactdensity0}
	n(r)&=\frac{2 (\beta +2)}{\pi  \left(\beta -\beta  r^2+2\right)^2}, \\
	\psi(r)&=-\frac{2}{\beta} \log \frac{\beta +2}{\beta(1 - r^2)+2},
\quad \psi(0)=0, \, \psi'(0)=0,\\
\psi(r)&=\frac{2 }{\beta }\log \left(1+ \frac{\beta(1 - r^2)}{2}\right), \quad \psi(1)=0, \, \psi'(0)=0 .
}
This solution is valid for $\beta>-2$. The corresponding stream function could be different depending on the boundary conditions. The vortex density Eq.~\eqref{exactdensity0} exhibits distinct behaviors in different parameter regimes. The vortices accumulate around the edge for $0<\beta$ while for $-2<\beta<0$ the vortices are center-concentrated (see Fig.\ref{f:exact0}). Note that in some literature, Eq.~\eqref{chiral1} does not have the prefactor $4\pi$ and hence the solution looks slightly different~\cite{Menon2008, Vishal2021}. 

\begin{figure}[h!t]
	\includegraphics[width=2.8in]{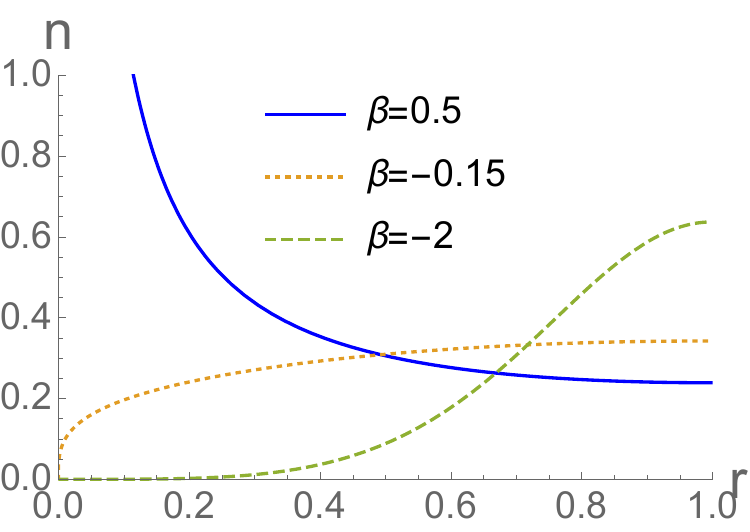}
	\vspace{0.1cm}
	\caption{Typical profiles of the vortex density distribution described by Eq.~\eqref{exactdensity} in three distinct parameter regimes: $0<\beta<1$, $-1/2<\beta<0$ and $\beta<-1/2$. 
	}
	\label{f:exactdensity1}
\end{figure}

Distinct from Eq.~\eqref{exactdensity0}, our  finding is Eq.~\eqref{densitynegative}:
\bea{\label{exactdensity}
	n (r)&=\frac{2(2/\beta-1)(1-\beta)^2 r^{-2\beta}}{\pi \beta (r^{-2\beta+2}+1-2/\beta)^2},\\
	\psi(r)&=-\frac{2}{\beta} \log \left[\frac{2 (\beta -1) r^{\beta }}{(\beta -2) r^{2 \beta }+\beta  r^2}\right],
}
with boundary conditions
\bea{\psi(1)=0, \quad  \psi'(1)=0.}
The solution Eq.~\eqref{exactdensity} holds for $\beta<1$ and $\beta \neq 0$. If requiring that $n'(r=0)$ is finite, $\beta<-1/2$.  For $0<\beta<1$, Eq.~\eqref{exactdensity} shows center-concentrated distribution and $n(r\rightarrow 0)\rightarrow \infty$. For $-1/2<\beta<0$, the vortex density is singular at origin, namely $n'(r\rightarrow 0) \rightarrow \infty$.  Vortices distribute around the edge for $\beta<-1/2$.  In contrast to the known exact solution Eq.~\eqref{exactdensity0}, the distribution Eq.~\eqref{exactdensity} is peaked on the boundary at negative temperature and  is maximized at  origin at positive temperature.  Figure \ref{f:exactdensity1} shows typical behaviors of the vortex density in these parameter regimes. 

\section { Conclusions }
Axis-symmetric clustered vortex  states for a neutral vortex system confined to a disc are investigated.  Combining the perturbation theory near the uniform state and  asymptotic analysis at high energies, we  find the parameter regime for which the rotation free states are supported.   At large energies,  the  distributions of positive vortices and negative vortices are well-separated and  the  edge-concentrated part provides a new exact vortex density distribution for the corresponding chiral vortex system.

The onset of a non-axisymmetric vortex cluster  in chiral vortex systems
appears to proceed via a second-order phase transition~\cite{SmithPRL, SmithONeil}.
It would be interesting to investigate possible  non-axisymmetric states for neutral systems carrying finite angular momentum. Thanks to the recent experimental advances~\cite{Gauthier2019, Johnstone2019, Mattprx},  our work would motivate  experimentally investigating  axis-symmetric  clustered phases in a homogeneous  Bose-Einstein condensate  trapped in cylindrically symmetric potentials.  Due to the presence of  conservation of angular momentum, axis-symmetric  clustered phases  are expected  to have longer life time than the giant vortex dipole state~\cite{Gauthier2019}.

\section*{Acknowledgement}

We acknowledge J. Nian, T. P. Billam, M. T. Reeves and A. S. Bradley for useful discussions.  X.Y. acknowledges support from the National Natural Science Foundation of China (Grant No. 12175215), the National Key Research and Development Program of China (Grant No. 2022YFA 1405300) and  NSAF (Grant No. U1930403).



%

\end{document}